# Stellar Intensity Interferometry with Air Cherenkov Telescope arrays

S. Le Bohec[1], M. Daniel[2], W.J. de Wit[2], J.A. Hinton[2], E. Jose[1], J.A. Holder[3], J. Smith[1], and R.J. White[2]

1: Department of Physics, University of Utah, 115 S 1400 E, Salt Lake City, UT 84112-0830, USA
2: School of Physics and Astronomy, University of Leeds, Leeds, LS2 9JT, UK
3: Department of Physics and Astronomy, University of Delaware, Newark, Delaware, USA

**Abstract.** The present generation of ground-based Very High Energy (VHE) gamma-ray observatories consist of arrays of up to four large (> 12m diameter) light collectors quite similar to those used by R. Hanbury Brown to measure stellar diameters by Intensity Interferometry in the late 60's. VHE gamma-ray observatories to be constructed over the coming decade will involve several tens of telescopes of similar or greater sizes. Used as intensity interferometers, they will provide hundreds of independent baselines. Now is the right time to re-assess the potential of intensity interferometry so that it can be taken into consideration in the design of these large facilities.

## STELLAR INTENSITY INTERFEROMETRY

The quantum theory behind Intensity Interferometry (I.I.) is due to Roy Glauber (6). The technique had earlier been understood in classical terms by R. Hanbury Brown and R. Twiss (9). In addition to the shot noise, the beating of Fourier components in starlight is responsible for an extra intensity fluctuation component, the wave noise. Because of this, fluctuations $\Delta i_1$ and $\Delta i_2$ recorded by two telescopes separated by a distance $d$ and pointed at a star, display a degree of correlation $c(d) = \frac{\langle \Delta i_1 \cdot \Delta i_2 \rangle}{\langle i_1 \rangle \langle i_2 \rangle}$ which is a measurement of $|\gamma(d)|^2$, the squared degree of coherence. $\gamma(d)$ is related by a Fourier transform to the light source profile along the telescope baseline direction (see Figure 1). This technique was exploited until 1971 in the Narrabri interferometer (8) for measuring 32 stellar diameters, some of them as small as 0.4 milli-arc-seconds (mas). These were the first direct measurements of stellar diameters since A.A.Michelson and F.G.Pease measured 5 stars with their 20 foot optical interferometer on Mount Wilson in 1920 (15), more than 40 years earlier. Since the Narrabri measurements, intensity interferometry has been largely ignored in favor of Michelson interferometry.

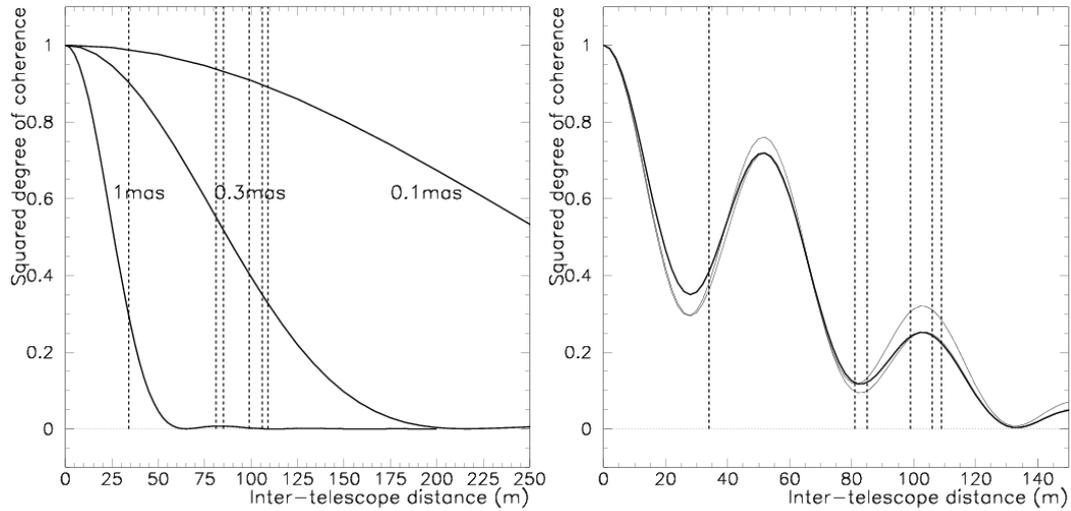

**FIGURE 1.** The squared degree of coherence as measured with a two telescope intensity interferometer is shown as a function of the inter-telescope distance for a wavelength of 440nm. The dashed vertical lines indicate the base lines available in the VERITAS array during observations at zenith (see Figure 6, left). On the left, the curves correspond to stars modeled as uniform disks of various diameters. On the right the thick curve corresponds to the Spica binary system during maximal separation along telescopes baseline. Each star is modeled as a uniform disk. Tidal effects and mutual irradiation effects were not included. The position of the minimum and maximum carries information about the distance between the two stars. One of the thin curves corresponds to the same system in which the diameter of one star has been reduced by 10%. The other thin curve results from a 10% increase in the diameter of the other star.

Michelson interferometry, relying on the measurement of the visibility of fringes, presents the advantage of allowing the measurement of bright stars with relatively small telescopes. I.I., on the contrary, requires great quantities of light and hence very large collectors. Michelson interferometry also presents much more stringent technical requirements. The path lengths of the light must be controlled to a fraction of the wave-length requiring atmospheric turbulences to be actively compensated for. With I.I., path lengths must only be controlled to a fraction of the coherence time: a few centimeters for a *GHz* bandwidth system. Light collectors in an intensity interferometer can be relatively crude and inexpensive for their size. Selecting the shortest wave-length of the visible spectrum does not affect the requirements for intensity interferometry. In a Michelson interferometer, delay lines consist of high precision optics and mechanics while in an intensity interferometer, simple electronic delays can be used. Also, with I.I., once intensity fluctuations have been converted into electronics signals, they can be duplicated indefinitely to be used in any number of interferometric pairs or multiplets without loss of sensitivity. It was demonstrated theoretically first (5) and then experimentally (23) that even with I.I., phase information can be recovered with a system consisting of three or more telescopes, by analyzing third and higher order correlation functions. I.I., as with Michelson interferometry, allows model-independent reconstruction of the object image (4, 27, 14, 10). Conversely, it was recently shown that if a model can be used, higher order correlations can be used to improve the sensitivity (17,18).

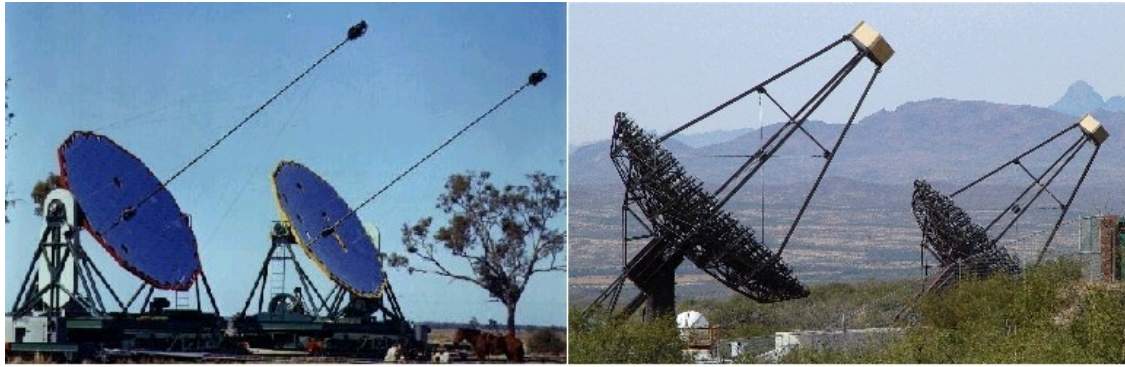

**FIGURE 2.** The two telescopes used in the Narrabri Interferometer are shown on the left. The two first VERITAS telescopes shown on the right could also be used as an interferometric pair of receivers.

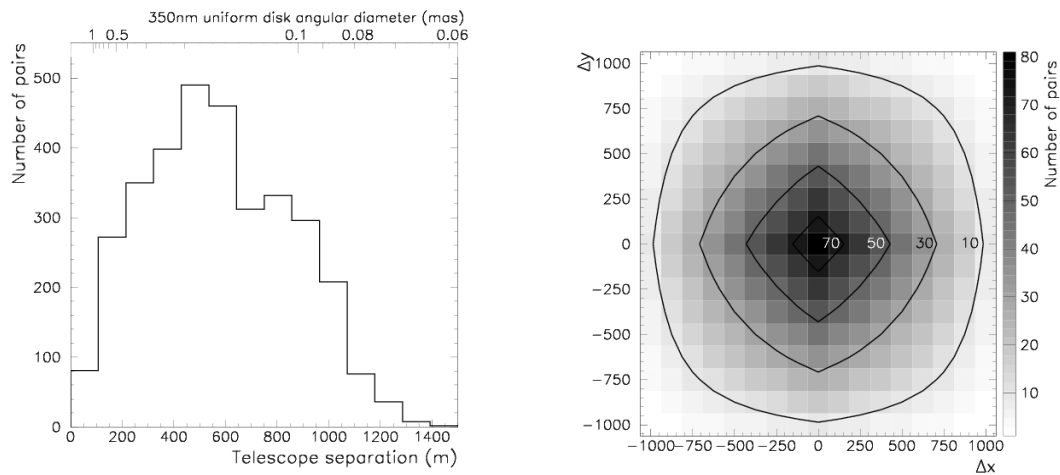

**FIGURE 3.** The distribution of baselines in a 1km² square grid array of 81 125m spaced telescopes is shown on the left. The upper scale indicates the first cancellation base line for a uniform disk observed at 350nm. The two dimensional baseline distribution is shown on the right.

# AIR CHERENKOV TELESCOPE ARRAYS

Figure 2 illustrates the similarities between the receivers in the Narrabri Intensity Interferometer and in VERITAS, a modern Air Cherenkov Telescope (ACT) array used for gamma-ray astronomy in the energy range above $100\,\text{GeV}$. In fact, the Narrabri telescopes were also used for gamma-ray observations (7). The ACT technique relies on the fact that high energy particles, including gamma rays, when entering the atmosphere, initiate extensive "air showers" of high energy secondary particles, some charged and radiating Cherenkov light. When it reaches the ground, the Cherenkov flash from a shower is very brief: only a few nanoseconds. It is also very faint, $\sim 10\,\text{photons}\cdot\text{m}^{-2}$ at $100\,\text{GeV}$ and large light collectors equipped with fast electronics are necessary. In order to record stereoscopic views of each shower, ACTs are now generally used in arrays of 2 to 4 units with typically 100m intertelescope distances, to match the extent of the Cherenkov light pool (28). Future projects such as CTA or AGIS aim at improving the gamma ray flux sensitivity by increasing to several square kilometers the area covered by the array with several tens of telescopes. Figure 3 shows the baselines that would be offered by an array which, when used as an interferometer, would allow to probe angular scales from a few mas to a few hundredths of a mas.

## A MODERN INTENSITY INTERFEROMETER

The width of the used optical band has no effect on I.I. sensitivity. A ~10nm band is convenient and easy to obtain. The light from the mirror must be collimated before being filtered and then concentrated on one or more photo-detectors. Several optical bands could also be used simultaneously. This can be achieved with lenses, interferometric filters or gratings and beam splitters (see Figure 4). The photo-detector signals must then be collected where the correlator is located. Photo-detection, signal communication and processing in the correlator can all benefit from technological developments since the time of the Narrabri interferometer.

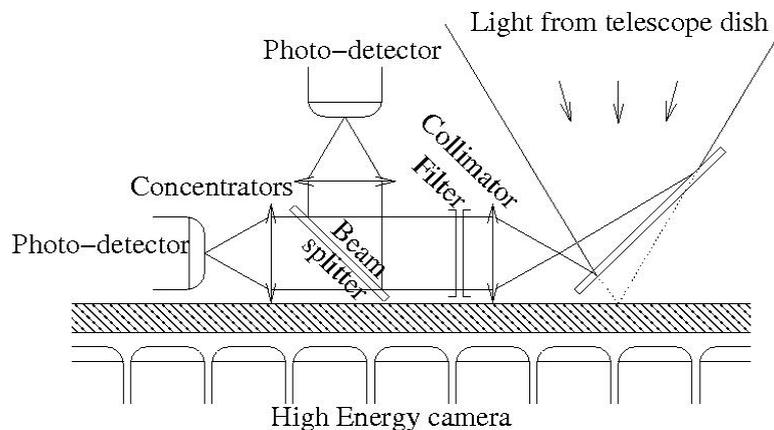

**FIGURE 4.** If I.I. capability is not implemented directly in the high energy camera, it could be done on a plate to be installed in front of the high energy camera for interferometric observations. Having more than one channel per telescope and optical band allows zero baseline monitoring for calibration purposes. See text for details regarding the optical design.

Photo-multipliers offer the possibility of a high bandwidth and high gain but they suffer from a relatively low quantum efficiency, typically less than 30% while the sensitivity of an intensity interferometer is proportional to the quantum efficiency. With quantum efficiencies close to 60% Geiger-APDs are very promising. They are now becoming available with high signal bandwidth and gains from $10^5$ to $10^6$. The Geiger mode of operation also provides an excess noise much smaller than in standard photo-multipliers. These photo-detectors are being considered for ACT high energy imaging cameras (19) and they should be considered for I.I. projects as well.

Under low illumination level, G-APDs as well as photomultiplier tubes can be used to produce a train of pulses corresponding to the photon streams at the two telescopes to be correlated. This type of approach is being pursued by Dainis Dravins (3). Alternatively, it is possible to use FADCs to digitize and record the intensity fluctuations (13). In both cases, the correlator can now easily be implemented in Field Programmable Gate Arrays (FPGA). FPGAs can be used with clocking rates of several hundred MHz.

The sensitivity of an intensity interferometer is proportional to the square root of the signal bandwidth. With bandwidths reaching 1GHz and baselines over 100m, coaxial cables are excluded. The transfer of signals to the correlator can be achieved using high bit-rate optical fibers. Optical fibers also present the advantage of eliminating the need for a common ground over the large distances separating the receivers.

## DEVELOPMENTS IN UTAH

These technological developments outlined in the previous section and the perspective of future large arrays of large light collectors, suggests that now might be the time to revisit stellar intensity interferometry. The group in Utah is pursuing an approach that consists of digitizing the intensity fluctuations so as to have them correlated in a central FPGA. A first correlator prototype has been constructed (see Figure 5, left). In this two-input correlator the signals are digitized into 12bits at a 200MHz rate. A FPGA collects the samples from both FADCs. After a 5ns resolution programmable delay (also implemented in the FPGA), the samples are multiplied and accumulated in a register. In order to obtain finer control of the relative timing of the signals, the FPGA programs 0.6ns steps analogue delays on each input. The FPGA also controls the analogue phase modulation. The demodulation is obtained by alternately adding and subtracting the sample's product, according to the states of the analogue phase switches (see Figure 5, right). This allows for the effects of slow analogue drifting, offsets and FADC pedestals to be canceled out. The FPGA is also programmed to include a processor that handles serial communication and data-transfer with a computer. This prototype is now fully working and tests in the laboratory with an artificial source and two photo-multipliers are under preparation.

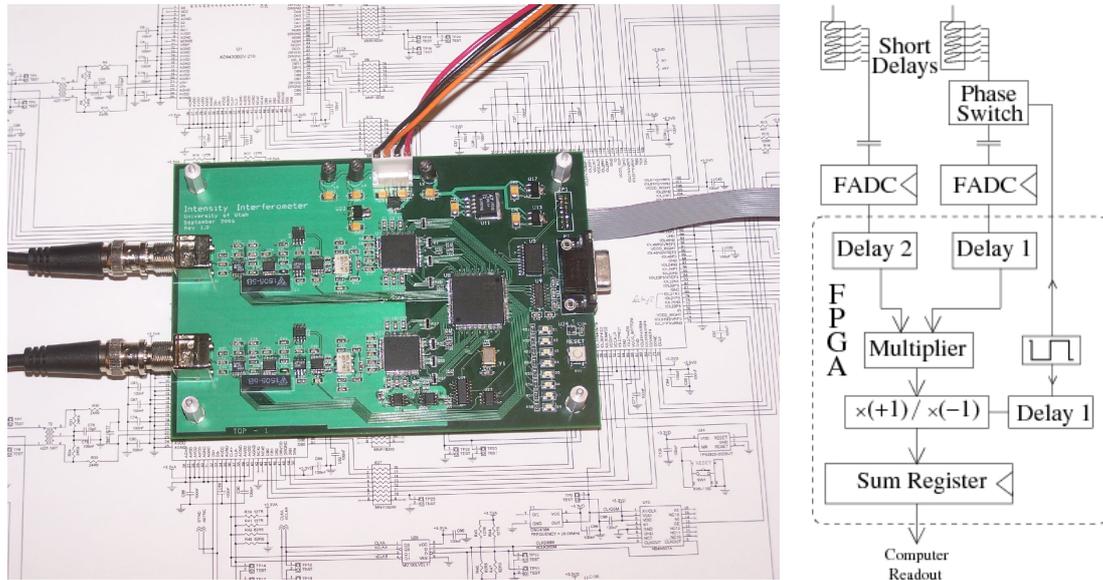

**FIGURE 5.** The digital correlator prototype is shown on the left. On the left of the board are the two inputs with programmable gain amplifier, fine delay and phase switch. The two chips in the center are the 200MHz FADCs. The big chip on the right is the FPGA programmed to process the signals and communicate with the computer. On the right is a system functional schematic, see text for details.

Such a correlator could be used with a pair of telescopes for stellar diameter measurements; however, its design requires the analogue signals to be sent from the telescopes to the correlator unit where they are digitized. This is certainly possible with conventional cables as long as the inter-telescope distance and signal bandwidth remain small enough. Alternatively it is possible to use analogue fibers (see next section) In future versions, the digitization could also be done locally at each telescope. A clock signal would be distributed by optical fiber from the central station to each telescope, along with analogue phase switching and fine delay control signals.

The authors have access to the arrays of 12m telescopes in H.E.S.S. and VERITAS (Figure 6, left) and this system, as well as future improved versions, could be tested on these instruments. In fact Dainis Dravins and Stephan LeBohec are preparing a series of tests with the VERITAS telescopes. However the best procedures and technical difficulties have not been all fully identified yet. During this development phase, a dedicated test-bed instrument would present a strong advantage. For this reason, two such telescopes, 3m in diameter (see Figure 6, right), are being installed in Grantsville, Utah at the SeaBase diving center (24). On an east-west 23m baseline, this pair of telescopes will allow the testing of various detectors, electronics and correlators for I.I. on first magnitude stars. This facility will become operational during winter 2007-2008 and first tests on the sky with the prototype correlator are anticipated for spring 2008.

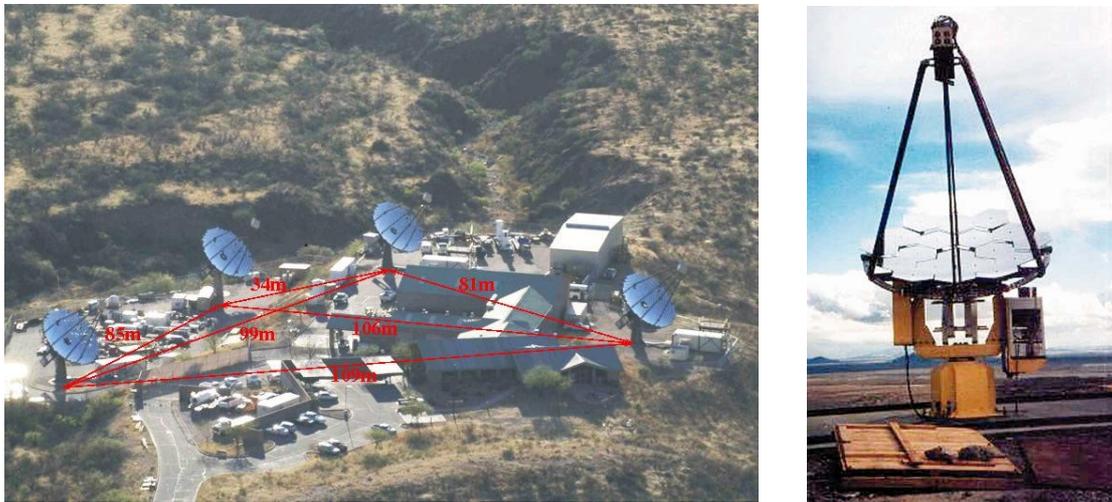

**FIGURE 6.** The VERITAS array, shown on the left, consists of 4 telescopes, 12m in diameter with inter-telescope baselines ranging from 34m to 109m. Before using these instruments, two 3m telescopes such as the one shown on the right are being installed in Utah for prototyping.

## DEVELOPMENTS IN LEEDS

The correlator under development in Utah requires the distribution of short fast pulses, typically of 3-5 ns FWHM and 1-2 ns rise time between the telescopes and the correlator unit over distances on the order of ~100m. Coaxial cable attenuates the pulse amplitude by ~25% and imparts a dispersion of over 50% into the FWHM of a typical pulse and is therefore not suitable for such an application. Conversely an optical link experiences no crosstalk or electromagnetic pickup between channels, no grounding problems and is immune to lightening strikes. Vertical Cavity Surface Emitting Lasers (VCSELs) are low power, low cost, readily available laser diodes capable of transmitting nanosecond rise-time signals over hundreds of meters of optical fibre with virtually no attenuation and are therefore a natural choice for use in a high-speed optical link. Typically employed in digital systems the use of VCSELs in analogue transmission schemes is novel, but not unique (21, 22). Sporadic changes in amplitude of the transmitted pulses were observed in these previous attempts and attributed to laser-mode hopping. Recent improvements in the VCSEL manufacturing technique suggest it is now timely to consider the next generation of analogue transmission scheme.

A prototype analogue transmission scheme has been developed in Leeds. A single channel transmitter (Figure 8, left) converts an input electrical pulse into an optical form using a Zarlink ZL60052 VCSEL. The VCSEL output is coupled into a 62.5 micrometer core, multi-mode, optical fibre via an E2000 connector, transported from the telescope to the correlator unit and converted back to an analogue electrical form using a photodiode based receiver. The linearity of the optical link is shown on the left side of Figure 7 together with the measured and calculated RMS noise (20). The link is linear to within 10% across a dynamic range of 61 dB. The bandwidth of the transmitter measured over 100 m of optical fibre is 470 MHz, as shown on the right side of Figure 7, although the bandwidth of the link is currently limited to ~250 MHz by the receiver. The frequency response of coaxial cable is also included in Figure 7, and is considerably less than that of the optical link.

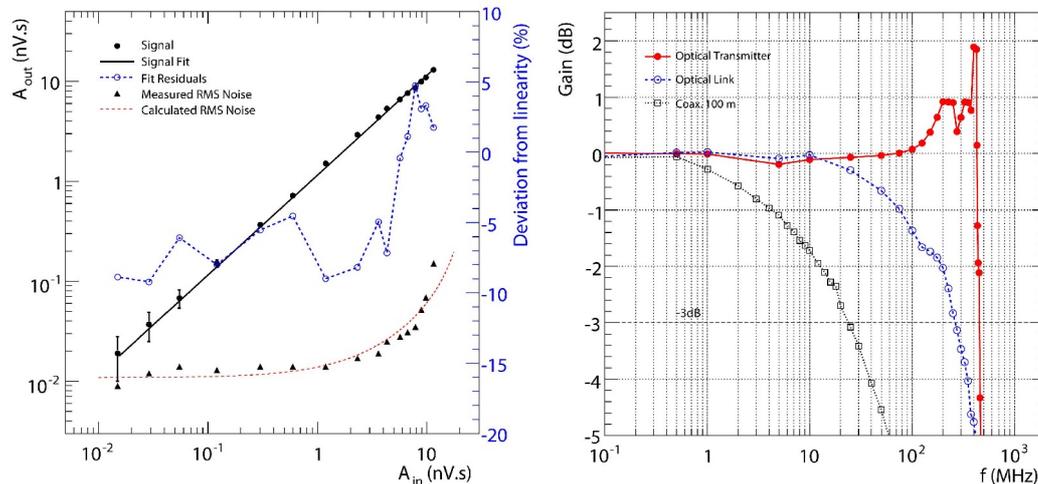

**FIGURE 7.** The measured area of the output pulse from the analogue optical link as a function of the measured input pulse area (filled circles) with a linear fit (solid line) is shown on the left. The fit residuals (open circles) indicate the linearity of the optical link. The measured RMS noise as a function of input pulse area is given by the triangular symbols, and the calculated RMS noise is given by the dashed continuous line. The gain of the optical link transmitter (filled circles), optical link as a whole (open circles) and 100 m of coaxial cable (open squares) as a function of sinusoidal input signal frequency are shown on the right.

In future versions of the correlator, the digitization could be performed locally at each telescope, requiring a clock signal to be distributed from the central station to each telescope, along with analogue phase switching and fine delay control signals. A digital asynchronous transceiver (DAT), developed in Leeds for high-speed signal distribution across optical fibre may provide one solution. In their current incarnation the DAT modules consist of two single-width, 6U high VME modules, denoted DAT-TX and DAT-RX as shown on the right side of Figure 8 and are used for the distribution of trigger signals, event numbers and housekeeping information within VERITAS (29). The conversion of electrical signals to optical signals is achieved using the Infineon 1.25 Gbit/s parallel optical link (PAROLI) consisting of a small form factor, 12 channel, 850 nm VCSEL driven transmitter and PIN diode array based receiver. To reach edge speeds of under a nanosecond the PAROLI devices make use of LVDS differential signalling. The PAROLI has a range of around 800 m that puts an upper limit on the distance between DAT modules and therefore telescopes.

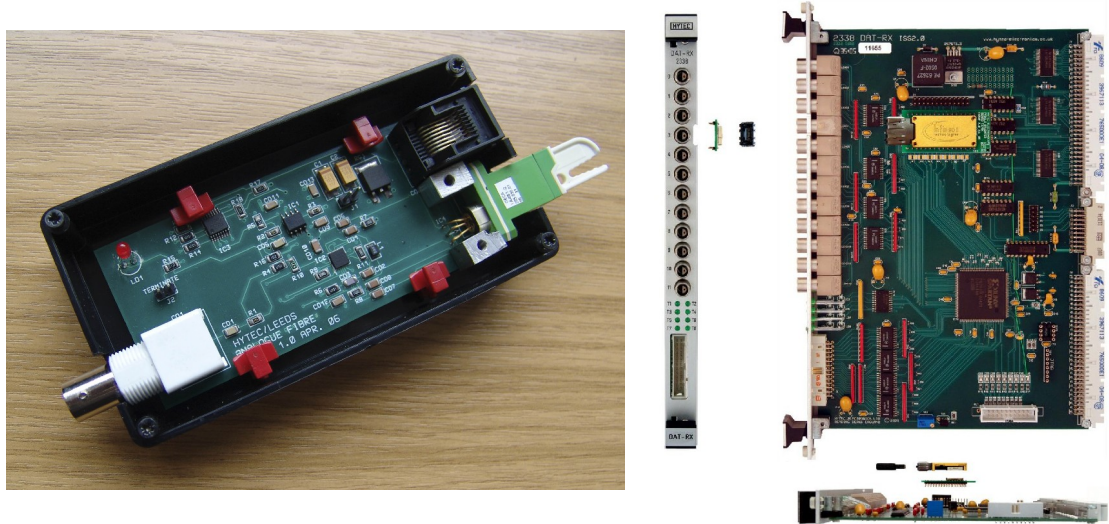

**FIGURE 8.** The single channel analogue optical transmitter prototype is shown on the left. On the right is The DAT-RX receiver module (DAT-TX left, DAT-RX right) with the PAROLI shown towards the top of the board. Signals are input via the front panel of the DAT-TX and presented at the front panel of the DAT-RX.

As a matter of laser safety, a given PAROLI transmitter channel will become disabled if a signal exceeds a duty cycle (DC%) of 57% within 1 microsecond. Input data violating this condition cannot be sent to the PAROLI directly, and must first be modulated to the DC% requirement in a recoverable manner. This is achieved using XOR encoding and decoding embedded within onboard Xilinx Spartan 3 FPGAs. The use of combinatorial operations avoids the dead time associated with sequential logic, and the recovered data has a RMS jitter of around 250 ps. The maximum data rate transmittable in this way is 200 MHz. Clock signals of 50% DC% at frequencies above 1 MHz do not require XOR encoding to satisfy the DC% condition and may be transmitted directly by disabling the encoding scheme over VME interface, also embedded in the FPGAs. The maximum clock in this way is 1.25 GHz. The DAT technology is therefore suitable for high-speed clock and control signal distribution within I.I., although a further iteration of the modules is required to adapt the DAT-TX into a form usable with the camera of an ACT.

## PROSPECTS

The techniques outlined above allow the implementation of I.I. over virtually unlimited baselines. We expect that the prototypes under construction will soon experimentally demonstrate this. The important questions that remain are whether the science motivations exist for reviving a technique abandoned 40 years ago and what is the sensitivity that can be achieved by an intensity interferometer implemented on ACT arrays like VERITAS and HESS or with future larger scale projects.

In a simple model, the sensitivity of an intensity interferometer is given by $(S/N)_{RMS} = A \cdot \alpha \cdot n |\gamma|^2 \sqrt{\Delta f \cdot T/2}$ where $A$ is the light collection area, $\alpha$ the photo-detector quantum efficiency, $n$ the spectral density (photons per unit optical

bandwidth, unit area and unit time), $\gamma$ is the degree of coherence, $\Delta f$ the electronics signal bandwidth and $T$ the observation time. With $|\gamma|^2=0.5$, $A=100\text{m}^2$ as in VERITAS or HESS and conservative parameters for the photo-detectors and electronics ($\Delta f=100\text{MHz}$ and $\alpha=25\%$), it seems that stars of visual magnitude 5 could be measured in 4 hours, providing a stellar diameter measurement with a precision of 3%(13). This can also be verified by simple scaling of the capabilities of the Narrabri interferometer. Fainter stars could be measured over longer observation time. This is, however, limited by the night sky background luminosity. The above sensitivity estimates assume the noise only consists of the Poisson fluctuation in the light from the measured star. The finite optical point-spread function of an ACT telescope integrates some of the surrounding night sky. With a point-spread function of $0.05°$ as in the VERITAS or HESS telescopes, non resolved ($|\gamma|=1$) stars of magnitude 9.6 are at the limit of observability. Using higher quantum efficiency detectors, larger telescopes or a higher bandwidth system will reduce the time necessary to reach this limit but will not improve the limitation itself. The only way to push this limitation is to improve the optical point-spread function. A very achievable point-spread function of $0.01°$ with 12m light collector would allow the measurement of stars with magnitudes up to 13, a possibility that could be taken into account for the design of future arrays.

I.I. will very likely never be used for measuring stars much fainter than magnitude 13, while Michelson interferometers can, in principle, be used for even fainter objects. Intensity interferometers can only be used for brighter stars; however, future large IACT arrays (see Figure 9, left) will make possible observations over a huge number of simultaneous baselines: a feat that is not even close to being achieved with amplitude interferometry (see Figure 3). ACT arrays do not usually operate while the Moon is visible, while I.I. with a narrow optical bandwidth can operate on sufficiently bright stars during all phases of the Moon. This competition-free access to telescope time makes long-term monitoring of variable objects very practical. The implementation of an intensity interferometer as a piggy-back system on existing large ACT arrays is extremely cheap and relatively simple, and could potentially greatly increase the scientific output of these expensive facilities. Additionally, intensity interferometers can operate with the same sensitivity over the entire visible and infra-red domains. A modern intensity interferometer could address the same topics as todays more conventional Michelson interferometers (2). Angular measurements are affected by limb darkening in a way that depends on the wavelength, so intensity interferometer measurements over a range of wavelengths could provide important information (16). Very long base-lines offer the possibility of measuring main sequence late-type stars or even Cepheids (see Figure 9, right) for which data is still missing (11,12). The measurements of the oblateness of rapid rotators (26) and the resolution of spectroscopic binaries (1) are also of interest. Selecting an emission line would also give access to the circumstellar environment (25). Finally, I.I. with large ACT arrays offers the opportunity to image nearby stars and systems of stars with unprecedented resolutions of a few hundredth of mas.

# ACKNOWLEDGMENTS

This work is supported by the University of Utah Funding Seed Grant program. SLB is grateful to Masahiro Teshima as well as to the Cosmic Ray group at the University of Utah for having given him access to the telescopes that will be used for the first tests. SLB is also grateful to Linda Nelson and George Sanders at SeaBase for their help and hospitality without which the deployment of the test telescopes would not be possible. Figure 3 results from the computational assistance of Louis LeBohec.

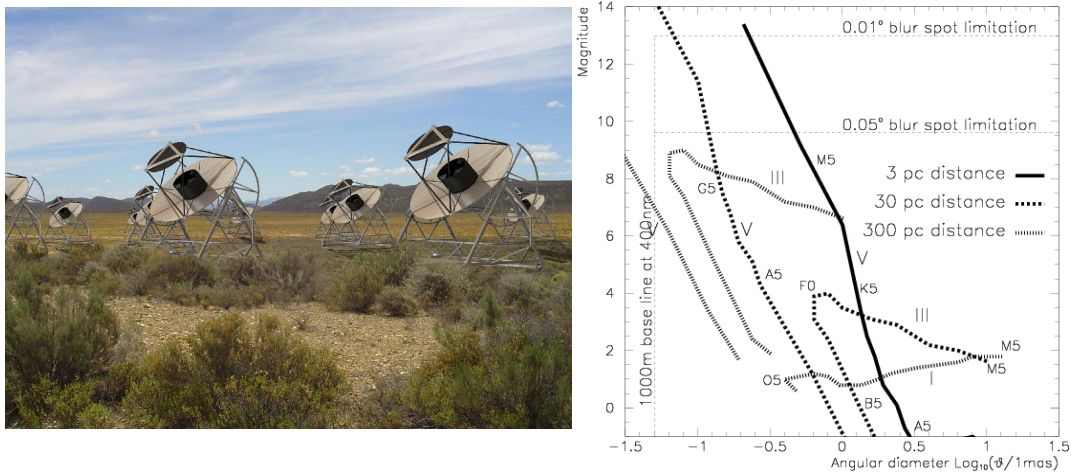

**FIGURE 9.** Ground based gamma ray observatory projects like AGIS as shown on the left (courtesy of J.Buckley and V.Guarino), could be used to measure and even image all stars in the main sequence within a 3pc distance all stars in the giant branch as far as 30pc and super giants to distances greater than 300pc.